\documentclass[twocolumn,showpacs,remark,amsmath,amssymb,pre]{revtex4}

\usepackage{amsmath}
\usepackage[dvips] {rotating}
\usepackage{epsf}
\usepackage{psfig}
\usepackage{epsfig}
\usepackage{graphicx}   % Include figure files
\usepackage{dcolumn}   % Align table columns on decimal point
\usepackage{bm}            % bold math
\usepackage{textcomp}
\usepackage{psfrag}
\usepackage[mathscr]{eucal}

\newlength{\colwidth}
\begin {document}
%
%\draft
%
% ******************************************************************************
%
\title{Stripe--hexagon competition in forced pattern forming systems with broken up-down symmetry}
%
%
% ******************************************************************************
%
\author{R. Peter}
\author{M. Hilt}
\author{F. Ziebert}
\author{J. Bammert} 
\author{C. Erlenk\"amper}
\author{N. Lorscheid}
\author{C. Weitenberg}
\author{A. Winter}
\author{M. Hammele}
\author{W. Zimmermann}

%
% ******************************************************************************
%
\affiliation{Theoretische Physik, Universit\"at des Saarlandes,
           D-66041 Saarbr\"ucken, Germany
}
\date{October 21, 2004}

%
% ******************************************************************************
%   Abstract
% ******************************************************************************
%
\begin{abstract}
We investigate the  response of two-dimensional 
pattern forming systems
with a broken up--down symmetry, such as chemical reactions, to spatially
resonant forcing and propose related experiments. 
The nonlinear behavior immediately  above 
threshold is analyzed in terms of amplitude equations suggested 
for a $1:2$ and $1:1$ ratio between the wavelength of the spatial 
periodic forcing and the wavelength of the pattern of the respective
system. Both sets of coupled amplitude equations 
are  derived by a perturbative method  
from the Lengyel-Epstein model describing 
a chemical reaction  showing Turing patterns, which gives us
the opportunity to relate the generic response scenarios
to a specific pattern forming system. The nonlinear 
 competition between stripe patterns and distorted hexagons 
is explored and their range of existence, stability
and coexistence is determined. Whereas without 
modulations hexagonal patterns are always preferred  near onset of 
pattern formation, single mode solutions (stripes) 
are favored close to threshold for  modulation amplitudes beyond some
critical value. Hence distorted hexagons only occur in a finite range of 
the control parameter and their interval of existence shrinks to zero
with increasing values of the modulation amplitude. Furthermore depending
on the modulation amplitude the  transition between stripes  and distorted hexagons 
is either sub- or supercritical.
  
\end{abstract}
%******************************************************************************
% PACS
% ******************************************************************************
% ******************************************************************************
%
\maketitle
\section{Introduction}%%%%%%%%%%%%%%%%%%%%%%%%%%%%%%%%%%%%%%%%%%%%%%%%%%%%%%%%%%
%%%%%%%%%%%%%%%%%%%%%%%%%%%%%%%%%%%%%%%%%%%%%%%%%%%%%%%%%%%%%%%%%%%%%%%%%%%%%%%%

External periodic forcing provides a powerful tool to analyze
the response behavior of nonlinear pattern forming systems, allowing 
for instance the study of their inherently nonlinear mechanism of 
self organization. The early investigations of effects dealing with
spatially periodic forcing were devoted to pattern formation in hydrodynamic
systems followed by chemical systems \cite{Kelly:78.1,Lowe:83.1,Lowe:86.1,Coullet:86.2,Coullet:86.1,Rees:86.1,Coullet:89.1,Hartung:91.1,Zimmermann:91.2,Zimmermann:93.3,Zimmermann:94.4,Zimmermann:96.1,Zimmermann:96.2,Zimmermann:96.3,Zimmermann:96.7,Stroock:2003.1,Zimmermann:1998.1,Kai:1999.1,Neubecker:2002.1,Epstein:2001.1,Epstein:2003.1,Ribotta:2003.1}. The situation is comparable for temporal forcing of patterns 
\cite{Hohenberg:84.1,Hohenberg:85.1,Riecke:88.1,Walgraef:88.1,Rehberg:88.1,Coullet:92.3,Riecke:92.2,Schatz:2000.2,Swinney:2000.1,Swinney:2000.2} whereas recent investigations 
on spatiotemporal forcing have mainly been motivated by chemical reactions
\cite{Zimmermann:2002.2,Sagues:2003.1,Kramer:2004.1,Schuler:2004.1}.

In thermal convection, when deviations from the Boussinesq approximation
come into play  \cite{Busse:67.1,Busse:89.2}, as well as in several 
chemical  reactions  of the Turing type \cite{Turing:1952.1,Murray:89},
the up--down symmetry of the fields 
describing the patterns, $u({\bf r},t) \to -u({\bf r},t)$, is broken. In such two--dimensional extended 
systems rotationally symmetric in a plane, hexagons are the generic pattern 
close to threshold. In systems with a strong axial anisotropy, such as in
nematic liquid crystals, this rotational symmetry  is broken 
\cite{deGennes:93,Zimmermann:89.4,Kramer:95.1,Kramer:96} and hexagonal 
convection  patterns do not occur. Thus the interplay between different
broken symmetries, such as the broken up--down symmetry with a weak anisotropy,
leads to an interesting competition between stripe and hexagonal patterns 
as has been shown recently  \cite{Zimmermann:1997.1,Ackemann:1999.1,Friedrichs:2002.1}.

Which kind of effects may be expected if besides the up-down symmetry 
the translational symmetry is simultaneously broken by periodic modulation 
in one spatial direction? As the direction of the modulation wave-number 
defines a preferred direction the system becomes anisotropic.
In the following we present the symmetry adapted 
amplitude equations and  analyze the related hexagon-stripe competition in
terms of these equations. In thermal convection this symmetry breaking may be
achieved by modulating the container boundaries 
or the applied temperature difference
periodically in one direction
 \cite{Kelly:78.1,Hartung:91.1,Zimmermann:96.3}.
In chemical reactions however, the forcing  is introduced by a spatially modulated
illumination of the system \cite{Epstein:2001.1,Epstein:2003.2}.
The specific manifestation of the modulation--induced symmetry breaking 
in generic amplitude equations depends on the ratio between the modulation wave 
number $k_m$ and the wave number $k_c$ of the respective pattern.
We will study two cases of resonant modulation: $k_m=2k_c$ and $k_m=k_c$.

Exemplarily we will focus in this work on the Lengyel--Epstein model for a
chemical reaction. This  model, introduced in  Sec.~\ref{LEM},
displays in some parameter range a  Turing instability of the homogeneous 
basic state as will be shown in  Sec.~\ref{LEMT}. Immediately above the threshold 
of a supercritical bifurcation the amplitudes of the unstable  pattern forming  
modes are still small enough to apply the powerful perturbational technique of 
amplitude equations. In Sec.~\ref{ampleq} we derive the amplitude equations
from the given Lengyel--Epstein model in the limit of small spatially periodic
modulations of the illumination acting on the chemical reactions. We consider a resonant ratio of 
$1:1$ as well as of $1:2$ between the wavenumber of the Turing instability and the
modulation wavenumber. The amplitude equations obtained in each of these resonant cases
are generic and may be found for other systems having the same symmetry
properties as the considered Lengyel--Epstein model, for instance modulated convection systems. Without spatially periodic forcing it is known that the resulting 
amplitude equations favor hexagonal patterns close to threshold. 
In  Sec.~\ref{nonlinsol} we investigate how the spatial forcing modifies the 
transition scenario between stripe and hexagon--like patterns and how the initial 
hexagonal patterns get distorted. Analytical calculations are confirmed by
numerical simulations in Sec.~\ref{Simul}. Concluding remarks and an outlook on 
traveling stripe forcing within the scope of the underlying chemical reaction are
given in Sec.~\ref{conclu}.

%%%%%%%%%%%%%%%%%%%%%%%%%%%%%%%%%%%%%%%%%%%%%%%%
\section{The Lengyel-Epstein model}
\label{LEM}
%%%%%%%%%%%%%%%%%%%%%%%%%%%%%%%%%%%%%%%%%%%%%%%%
As a basic model of a chemical reaction we choose the Lengyel--Epstein model
as described in Ref.~\cite{Epstein:1990.1}. If the effect of illumination on
the chemical reaction is taken into account, as done in  Ref.~\cite{Epstein:1999.1},
the model may be described by the following equations
\begin{subequations}
\label{LE}
\begin{align}
\label{LE1}
\partial_t u &= a - c u -4 \frac{uv}{1+u^2}-\Phi +\Delta u\,,\\
\label{LE2}
\partial_t v&=\sigma\left( c u -  \frac{uv}{1+u^2} +\Phi + d\Delta v\right)\,.
\end{align}
\end{subequations}
In these equations the dimensionless concentrations of two relevant chemical
substances are labelled with $u({\bf r},t)$ and $v({\bf r},t)$, whereas $a,c,\sigma$
and $d$ denote
dimensionless parameters of the chemical system. Finally the effect of external
illumination is described by the field $\Phi({\bf r},t)$, which can be  
identified as the control parameter of the system and decomposed into a spatially
homogeneous and a spatially periodic part 
\cite{Epstein:2001.1,Epstein:2001.2,Epstein:2003.1}
\begin{align}
\label{contle}
\Phi({\bf r},t) = \phi + M({\bf r})\enspace.
\end{align}
For the second term of Eq.~(\ref{contle}) we assume a periodic dependence in one
spatial direction as described by
\begin{align}
\label{contmod}
M({\bf r})= 2 \tilde G \cos(k_m x)\,.
\end{align}
For a spatially homogeneous illumination Eqs.~(\ref{LE}) have a stationary and
homogeneous solution given by
\begin{align}
\label{L0}
 u_0 = \frac{a -5\phi}{5c} \,, \qquad 
v_0=\frac{a(1+u_0^2)}{5u_0}\,.
\end{align}
%

%%%%%%%%%%%%%%%%%%%%%%%%%%%%%%%%%%%%%%%%%%%%%%%%%%
\section{Threshold for pattern formation}
\label{LEMT}
%%%%%%%%%%%%%%%%%%%%%%%%%%%%%%%%%%%%%%%%%%%%%%%%%%

\begin{figure}[hbt]
\begin{center}
\psfrag{aa}{}
\psfrag{kc}{$k_c$}
\psfrag{k}{\large $k$}
\psfrag{sig}{$\lambda(k)$}
\psfrag{phi1    }{$\phi<\phi_c$}
\psfrag{phi2    }{$\phi=\phi_c$}
\psfrag{phi3    }{$\phi>\phi_c$}
\includegraphics [width=8cm] {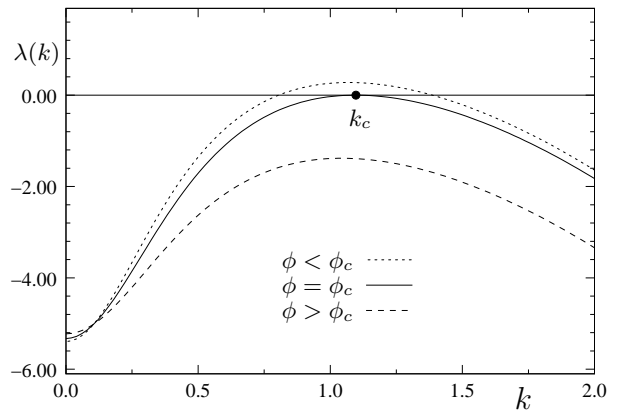}
%\vspace{3cm}
\end{center}
\caption{
The dispersion relation $\lambda(k)$ for a supercritical
value $\phi < \phi_c$ (dotted line) of the illumination
strength $\phi$, for a critical value $\phi=\phi_c$ (solid line)
and for a subcritical value $\phi>\phi_c$ (dashed line).
Parameters are $a=16,~c=1.0,~d=1.5, ~\sigma=301$.
}
\label{dispers}
\end{figure}

In a certain parameter range the spatially homogeneous state described by 
Eqs.~(\ref{L0}) becomes unstable against infinitesimal perturbations 
$u_1({\bf r},t)$ and $v_1({\bf r},t)$. In order to determine this interval the
basic state is separated from the inhomogeneous contributions to both fields 
by the ansatz 
\begin{subequations}
\begin{eqnarray}
 u({\bf r},t)&=&u_0 +u_1({\bf r},t)\enspace,\\
v({\bf r},t)&=&v_0 +v_1({\bf r},t)\enspace.
\end{eqnarray}
\end{subequations}
In a next step the linerization of the basic Eqs.~(\ref{LE}) with respect to 
those small contributions $u_1({\bf r},t)$ and $v_1({\bf r},t)$ yields linear 
equations with constant coefficients, which may be solved by the mode solutions 
\begin{eqnarray}
\label{modean}
u_1({\bf r},t)=  \bar{u}_1 e^{\lambda t + \dot{\imath}{\bf k}\cdot {\bf r}} \,, \qquad
v_1({\bf r},t)= \bar{v}_1  e^{\lambda t + \dot{\imath}{\bf k}\cdot {\bf r}}\,
\end{eqnarray}
and thus leads to a set of homogeneous linear equations for the amplitudes
$\bar{u}_1$ and $\bar{v}_1$.
From these equations the dispersion relation $\lambda(k)$ may be specified.
It turns out that a strong and homogeneous illumination of the chemical 
reaction suppresses inhomogeneous perturbations. This is illustrated by
Fig.~\ref{dispers} wherein the dispersion relation $\lambda(k)$ is plotted
for three different cases of the illumination strength $\phi$.
%for an 
%overcritical value of the control parameter $\phi < \phi_c$ (dotted line),
%for a critical one $\phi=\phi_c$ (solid line) 
%and for  an undercritical value $\phi> \phi_c$ (dashed line). 
%
%
% *************************************************************
%   Figure
% *************************************************************
%
%\fi
%.................................................................
%
%%
%
%
%
% *************************************************************
%   Figure
% *************************************************************
%
\begin{figure}[hbt]
\begin{center}
\psfrag{k}{\large $k$}
\psfrag{phi0}{\large $\phi_0$}
\includegraphics [width=8cm] {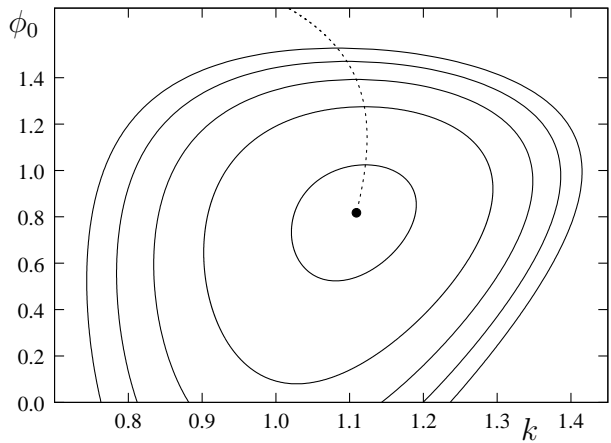}
%\vspace{3cm}
\end{center}
\caption{The neutral curve $\phi_0(k)$ is determined
numerically via the neutral stability condition in 
Eq.~(\ref{neucond}) for different values of the 
diffusion parameter $d=1.2, 1.3,...,1.6$ (from inside
to outside). The remaining parameters are 
$a=16,~ c=1.0,~ \sigma=301$ in all cases. The 
homogeneous state is unstable inside of each curve. The
dotted line shows $k_c$ as a function of $d$.
%respectively.
%Above 
%Inside the curve the homogeneous
%state described by Eq.~(\ref{L0} ) is stable and below unstable.
%The dashed curve marks the location of the maximum of the neutral
%curves, i.e. the critical wavenumber $k_c$ as given by Eq. 
%\eqref{kritwavenumber}. 
%Below a critical value $d=d_c$ the instability does not occur any
%more (see Fig. \ref{PhiKrit}), so this curve ends at this point.
}
\label{neutral}
\end{figure}
%\fi
%.................................................................
%
%

\begin{figure}[hbt]
\begin{center}
%
%\vspace{-5mm}
\psfrag{d}{\large $d$}
\psfrag{phi0}{\large $\phi_c$}
\includegraphics [width=8cm] {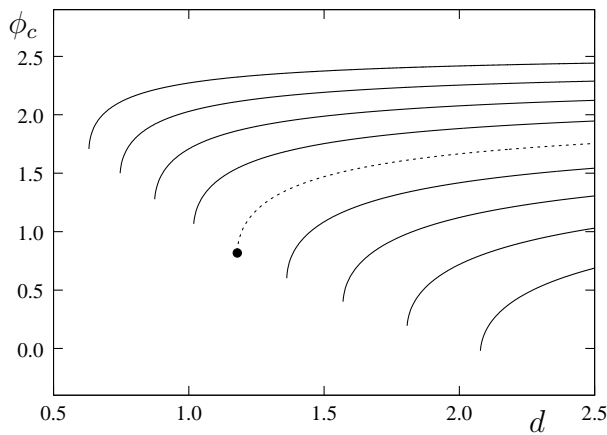}
\end{center}
%\vspace{-0.7cm}
%
\caption{
The critical illumination $\phi_c=\phi_0(k_c)$ is plotted 
as a function of the diffusion coefficient $d$ 
for different values of the parameter  $c$,
with 
$c=0.6, 0.7, 0.8, ...,1.4$ from top to bottom and 
$a=16,\sigma=301$. The dotted line corresponds to
that in Fig.~\ref{neutral}.}
\label{PhiKrit}
\end{figure}

From the {\it neutral stability condition}
\begin{eqnarray}
\label{neucond}
{\rm Re}(\lambda(k))=0, \quad 
\Rightarrow ~\phi_0(k)
\end{eqnarray}
the {\it neutral curve} $\phi_0(k)$  may be determined, which has to be
done numerically in our case. Some examples of characteristic neutral
curves are plotted in Fig.~\ref{neutral} for different values of the 
diffusion coefficient $d$.
For $d=1.2$ the homogeneous state is either stable for a small or for a strong 
illumination $\phi$ and is thus unstable only within a small island
enclosed by the corresponding neutral curve in the $\phi$-$k$-plane. However, 
for $d>1.4$ the homogeneous state remains only stable for strong illuminations.
In this case the neutral curve $\phi_0(k)$ takes its maximum at the wavenumber 
\begin{align}
k^2_c =\frac{1}{2} \left(\frac{4v_0(u_0^2-1)}{(1+u_0^2)^2} -c -  
\frac{u_0}{d(1+u_0^2)}\right)
\end{align}
and the value of the illumination at the maximum of the neutral curve defines the  
critical illumination strength $\phi_c=\phi_o(k_c)$. This critical value is plotted  
in Fig.~\ref{PhiKrit} as a function of the diffusion coefficient $d$ and for various 
values of the parameter $c$. The occurring Turing instability vanishes, when the inner
closed curve in Fig.~\ref{neutral} shrinks to zero, i.e. below $d_c=1.179$ for the 
parameter set chosen in this figure. Below this critical value $d_c$ the 
stationary homogeneous state becomes stable with respect to an oscillatory instability.

%
%
% *************************************************************
%   Figure
% *************************************************************
%
%\fi
%.................................................................
%

%%%%%%%%%%%%%%%%%%%%%%%%%%%%%%%%%%%%%%%%%%%%%%%%%%
\section{Amplitude equations and linear stability of  their 
solutions}
\label{ampleq}
%Amplitude equations and linear instability in  the 
%weakly nonlinear regime
%%%%%%%%%%%%%%%%%%%%%%%%%%%%%%%%%%%%%%%%%%%%%%%%%%
The amplitudes $\bar{u}_1$ and $\bar{v}_1$ of the linear modes
described by  Eq.~(\ref{modean}) as well as the amplitudes of 
any other pattern are restricted by the nonlinear terms proportional
to $u_1$ and $v_1$ in Eqs.~(\ref{LE}). In two--dimensional systems
with a broken up-down symmetry the common patterns close to threshold
are stripes, squares or hexagons \cite{CrossHO}.
Immediately beyond a supercritical bifurcation, where the amplitude
of an emerging pattern is still small, the concept of the so--called
amplitude equations is a very successful one to characterize the 
nonlinear behavior of patterns, as it is exemplified for several 
physical, chemical and biological systems in Refs.~\cite{Manneville:90,CrossHO}. 
The  amplitude equations are obtained by a perturbative method from 
the basic Eqs.~(\ref{LE}), whereby the coefficients occurring in these
amplitude equations reflect the dependence of the patterns on the 
parameters of the specific system.

%%%%%%%%%%%%%%%%%%%%%%%%%%%%%%%%%%%%%%%%%%%%%%%%%%
\subsection{Scheme of the derivation of amplitude equations}
\begin{figure}[hbt]
\begin{center}
%
%\vspace{-5mm}
\psfrag{d}{\small $d$}
\psfrag{tau0}{\hspace{-1.mm}\small $\tau_0$}
\psfrag{del}{\small $\delta$}
\psfrag{gam}{\small $\gamma$}
\psfrag{rho}{\small $\rho$}
\includegraphics [width=0.95\columnwidth] {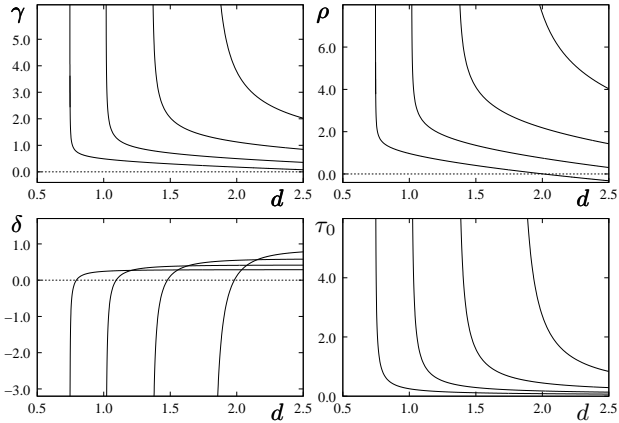}
%\includegraphics [width=0.48\columnwidth] {./figures/gamma.eps}
%\includegraphics [width=0.48\columnwidth] {./figures/rho.eps}
%\\[2mm]
%\includegraphics [width=0.48\columnwidth] {./figures/delta.eps}
%\includegraphics [width=0.48\columnwidth] {./figures/tau0.eps}
\end{center}
%\vspace{-0.7cm}
%
\caption{\label{coefamp}
The coefficients of the amplitude equations in 
Eq.~(\ref{amplihex}) are shown close to the $1:2$
resonance, i.e. $G_2=0$, as functions of the 
diffusion coefficient $d$ for different values of
the parameter $c$, from left to right 
$c=0.7, 0.9, 1.1, 1.3$.
}
\end{figure}

\begin{figure}[hbt]
\begin{center}
%
%\vspace{-5mm}
\psfrag{d}{\large $d$}
%\psfrag{rho_gam}{\hspace{-7mm}\Large $\bar \rho = {\rho \over \gamma}$}
\psfrag{rho_gam}{\Large $\bar \rho$}
\includegraphics [width=0.9\columnwidth] {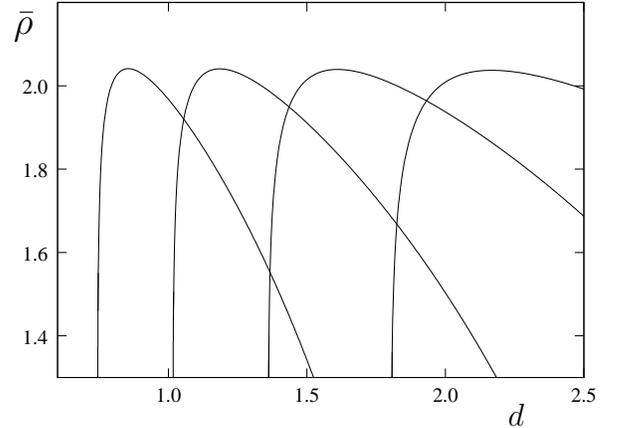}
\end{center}
%\vspace{-0.7cm}
%
\caption{\label{figrho}
The rescaled coefficient $\bar \rho = \rho / \gamma$
as a function of $d$ for various values of $c$, with 
$c=0.7, 0.9, 1.1, 1.3$ from left to right.
}
\end{figure}

%%%%%%%%%%%%%%%%%%%%%%%%%%%%%%%%%%%%%%%%%%%%%%%%%%
The small parameter used in the perturbative derivation of the amplitude
equations  is the relative distance to the threshold $\phi_c$
\begin{align}
\varepsilon = \frac{\phi_c-\phi}{\phi_c}\,.
\end{align}
Since the up-down symmetry for inhomogeneous field contributions is broken, 
i.e. $u_1,v_1  \not \to -u_1,-v_1$, we expect hexagonal patterns to occur
in some parameter ranges. Close to threshold hexagonal patterns are described  
by a superposition of three waves (stripe patterns) with wave vectors enclosing
an angle of about $120^0$ with respect to each other. 
If the two fields $u_1$
and $v_1$ are rewritten in terms of a vector $ {\bf w}_1 =(u_1, v_1)^T$, the 
generating field of hexagonal patterns may be  represented by  
\begin{align}
\label{hexansatz}
{\bf w}_1
%\left(\begin{array}{c} u_1\\v_1 \end{array} \right)
=& \left( \begin{array}{c} 1   \\ E_0 \end{array} \right)
\left(A_1 e^{i {\bf  k}_1 {\bf r}} + A_2 e^{i {\bf k}_2 {\bf r}} + A_3 e^{i {\bf k}_3 {\bf r}}\right) + c.c. \, ,
\end{align}
where the wave vectors  ${\bf k}_i$  ($i=1,2,3$) of the three underlying stripe
patterns read 
\begin{align}
{\bf k}_1     = k_c           \left(\begin{array}{c}  1 \\ 0 \end{array} \right)
\quad{\rm and}\quad
{\bf k}_{2,3} = {k_c \over 2} \left(\begin{array}{c} -1 \\ \pm  \sqrt{3} \end{array}\right)\,
\end{align}
and the second component $E_0$ of the eigenvector is given by
\begin{align}
E_0 = -\left(c+k_c^2\right)\frac{1+u_0^2}{4u_0}-\frac{v_0}{u_0}\frac{1-u_0^2}{1+u_0^2}\,.
\end{align}
The coupled amplitude equations deduced for the three  envelope functions
$A_i$ ($i=1,2,3$) are identical apart from the term evoked by the forcing 
which naturally differs whether we choose in Eq.~(\ref{contmod})
a ratio $1:2=k_c:k_m$ between the modulation wave number $k_c$ and the 
critical wave number $k_m$ or a ratio $1:1=k_c:k_m$. In the first case the 
modulation amplitude $G$ should vary as $G =G_1 \propto \varepsilon$, in the 
latter one as  $G=G_2 \propto \varepsilon^{3/2}$.\\  
Choosing the ansatz specified in Eq.~(\ref{hexansatz}) and expanding 
Eq.~(\ref{LE}) with respect to powers of the small parameter $\varepsilon$
as well as the small amplitudes $A_i$ one obtains the following set of nonlinear equations for the three 
amplitudes $A_i$   
\begin{subequations}
\label{amplihex}
\begin{align}
\tau_0 \partial_t A_1 =& \,\varepsilon\, A_1 + \delta\, A_2^\ast A_3^\ast +G_1 A_1^\ast +G_2\nonumber 
%+\alpha G_3
\\
 &\quad -\left( \gamma\, |A_1|^2 + \rho\, |A_2|^2 + \rho\, |A_3|^2 \right)A_1\,, \\
\tau_0 \partial_t A_2 =& \,\varepsilon\, A_2 + \delta\, A_3^\ast A_1^\ast  
%+G_3A_3^\ast
\nonumber \\
 &\quad -\left( \gamma\, |A_2|^2 + \rho\, |A_3|^2 + \rho\, |A_1|^2 \right)A_2\,, \\
\tau_0 \partial_t A_3 =& \,\varepsilon\, A_3 + \delta\, A_1^\ast A_2^\ast  
%+G_3A_2^\ast
\nonumber \\
 &\quad -\left( \gamma\, |A_3|^2 + \rho\, |A_1|^2 + \rho\, |A_2|^2 \right)A_3\,,
\end{align}
\end{subequations}
which are quite similar to those presented in Refs.~\cite{Ciliberto:90.1,Zimmermann:1997.1,Kramer:2004.1}. Depending on
which of the resonant cases mentioned above is considered either $G_1$
or $G_2$ is nonzero.
For further details on the derivation scheme of the amplitude equations
we refer to Ref.~\cite{Hilt:2004.1}, where likewise reaction diffusion
equations were studied in detail.\\
The analytical expressions of the coefficients $\tau_0,~ \delta, ~ \gamma,~ \rho$  
appearing in the amplitude equations (\ref{amplihex}) as functions of the parameters
of the basic equations (\ref{LE}) are rather lengthy. Instead of giving
their analytical forms we therefore plot them in Fig.~\ref{coefamp} as functions of 
the diffusion coefficient $d$ while assuming different values for $c$. 
%
%
% *************************************************************
%   Figure
% *************************************************************
%
%\fi
%.................................................................
%
Rescaling of the amplitudes $A_i$ $(i=1,2,3)$ allows to express time
as well as the coefficients of Eqs.~(\ref{amplihex}) as follows 
\begin{align}\label{scale}
\bar A &= \frac{\gamma}{\vert\delta\vert}\, A\, , \quad
&\eta &= \frac{\gamma}{\delta^2} ~\varepsilon\, , \quad
%&\bar B &= \frac{\gamma}{\vert\delta\vert}\,B \\
&\bar t &=  \frac{\delta^2}{\tau_0 \gamma}\,t\, ,\nonumber\\ 
\bar G_{1} &= \frac{ \gamma}{\delta^2}~G_{1}\, ,   
&\bar G_2 &= \frac{\gamma^2}{\vert\delta\vert^3}~G_2\, ,
& & &\nonumber\\
\bar \rho &=\frac{\rho}{\gamma} \, , 
&\bar \delta &= \frac{\delta}{\vert\delta\vert}
\end{align} 
and yields a simpler form of Eqs.~(\ref{amplihex}) 
\begin{subequations}
\label{amplihexs}
\begin{align}
 \partial_{\bar t} \bar A_1 =& \,\eta \, \bar A_1 + \bar \delta 
\bar  A_2^\ast \bar A_3^\ast +\bar G_1 \bar A_1^\ast +\bar G_2\nonumber 
%+\bar \alpha \bar G_3 
\\
 &\quad -\left( |\bar A_1|^2 + \bar \rho\, |\bar A_2|^2 
+ \bar \rho\, |\bar A_3|^2 \right)\bar A_1\,, \\
 \partial_{\bar t} \bar A_2 =& \,\eta \, \bar A_2 +\bar \delta
  \bar A_3^\ast \bar A_1^\ast  
%+\bar G_3 \bar A_3^\ast
\nonumber \\
 &\quad -\left(  |\bar A_2|^2 + \bar \rho\, |\bar A_3|^2 + \bar \rho\, |\bar A_1|^2 \right)\bar A_2\,, \\
 \partial_{\bar t} \bar A_3 =& \,\eta \bar A_3 +\bar \delta 
\, \bar A_1^\ast \bar A_2^\ast  
%+\bar G_3 \bar A_2^\ast
\nonumber \\
 &\quad -\left( |\bar A_3|^2 + \bar \rho\, |\bar A_1|^2 + \bar \rho\, |\bar A_2|^2 \right)\bar A_3\,
\end{align}
\end{subequations}
with $\bar \delta = \pm 1$. Choosing $\tau_0=\gamma=1$ and $\delta=\pm 1$  
in Eqs.~(\ref{amplihex}) reduces them 
to the same form as  Eqs.~(\ref{amplihexs})
%and the latter ones are 
which besides
the modulation amplitude $G_i$ $(i=1,2)$ and the control parameter $\eta$ 
only depend on the nonlinear coefficient $\bar \rho$. In most cases  
this coefficient takes values in the interval $\bar{\rho}\in[1,2]$ as shown in Fig.~\ref{figrho}, where $\bar\rho$ is plotted as a function of the diffusion 
coefficient $d$ for different values of $c$.
%
%
%
% *************************************************************
%   Figure
% *************************************************************
%
%\fi
%.................................................................
%
%%%%%%%%%%%%%%%%%%%%%%%%%%%%%%%%%%%%%%%%%%%%%%%%%%%%%%%%%%%
\subsection{Functional}
%%%%%%%%%%%%%%%%%%%%%%%%%%%%%%%%%%%%%%%%%%%%

The coupled nonlinear equations (\ref{amplihex}) may be considered
as a potential dynamics
\begin{align}
\tau_0 \partial_t A_i = - \frac{\delta {\cal F}}{\delta\! A_i^* }
\end{align}
with the functional
\begin{align}
\label{pothexa}
{\cal F} =&
 \sum_{i=1}^3 \left( \frac{\gamma}{2}\,|A_i|^4 -  \varepsilon\,|A_i|^2 \right)
+ \frac{\rho}{2} \sum_{j \neq i}^3 |A_i|^2\, |A_j|^2
\nonumber \\
& \quad -\frac{G_1}{2} 
\left(A_1^2+A_1^{\ast 2}\right)  -
%\left( 
G_2 
%+\alpha G_3\right) 
\left(A_1 +A_1^\ast\right)\nonumber \\
%& \quad -\frac{G_3}{2} 
%\left(A_2^2 +A_3^2+A_2^{*2}+A_3^{*2} \right)\nonumber \\
& \quad
-~ \delta\,\left( A_1 A_2 A_3 + A_1^* A_2^* A_3^* \right)\,.
\end{align}
This functional or its rescaled version yielding Eqs.~(\ref{amplihexs}) 
provides a useful criterion to decide which kind of pattern is most 
likely to occur in a given parameter range. Generally the pattern 
leading to the lowest value of the functional is preferred. Once again
either $G_1$ or $G_2$ is nonzero, corresponding to one of the considered
resonant cases.
%Since we investigate only one of the two resonance cases, namely the $2:1$ and $1:1$
%resonance,  always only one of the amplitude
%$G_i$ ($i=1,2$) is assumed to be different from zero. 
%%%%%%%%%%%%%%%%%%%%%%%%%%%%%%%%%%%%%%%%%%%%%%%%%%%
\subsection{Linear stability of hexagons and rolls}
\label{linstabs}
%%%%%%%%%%%%%%%%%%%%%%%%%%%%%%%%%%%%%%%%%%%%%%%%%%%
Since we only consider a spatially periodic modulation in $x$-direction,
the translational invariance in $y$-direction is preserved. This holds
for both resonant cases under investigation. Therefore two out of three
amplitudes, namely $A_2$ and $A_3$ are degenerated and the remaining two
stationary amplitudes, called $A$ and $B$, are determined by two of the three Eqs.~(\ref{amplihexs}). The linear stability of the stationary and nonlinear 
solutions is analyzed by superimposing small perturbations to the latter ones,
meaning we use the following ansatz:   
\begin{align}
\bar A_1= A + \mu_1(\bar t), \quad \bar A_{2,3}= B +\mu_{2,3}(\bar t)\enspace.
\end{align}
Accordingly the amplitude equations (\ref{amplihexs}) are
linearized with respect to the small perturbations $\mu_{i}$ $(i=1,2,3)$
which leads to the following set of equations
\begin{subequations}
\label{linstab}
\begin{align}
\partial_{\bar t} \mu_1 =& \,
\left(\eta -2 |A|^2 -2 \bar \rho |B|^2\right) \mu_1 
+\left( \bar G_1 -  A^2 \right) \mu_1^\ast 
\nonumber \\
 -&\bar \rho  A  B^\ast (\mu_2+\mu_3)  
+ \left( \bar \delta  B^\ast -\bar \rho   A  B
\right) (\mu_2^\ast+\mu_3^\ast)  \enspace ,\\
 \partial_{\bar t} \mu_2 =& \,
\left(\eta -( 2+\bar \rho) |B|^2-\bar \rho |A|^2 \right) \mu_2 - B^2 \mu_2^\ast 
\nonumber \\
&  -\bar \rho  A^\ast  B \mu_1  
+ \left( \bar \delta  B^\ast -\bar \rho   A  B
\right) \mu_1^\ast 
\nonumber \\
&  -\bar \rho |B|^2 \mu_3  
+ \left( \bar \delta  A^\ast -
\bar \rho   B^2 \right) \mu_3^\ast \enspace ,\\
 \partial_{\bar t} \mu_3 =& \,
\left(\eta -(2 +\bar \rho) |B|^2
-\bar \rho |A|^2\right) \mu_3 
-  B^2  \mu_3^\ast 
\nonumber \\
&  -\bar \rho  A^\ast  B \mu_1  + 
\left( \bar \delta  B^\ast -\bar \rho   A  B\right) \mu_1^\ast 
\nonumber \\
&  -\bar \rho |B|^2 \mu_2  
+ \left( \bar \delta  A^\ast -\bar \rho   B^2\right) \mu_2^\ast \,.
\end{align}
\end{subequations} 
These coupled equations may be transformed into a set of homogeneous
linear equations by means of an ansatz for the perturbations
\begin{align}\label{perans}
\mu_i = \nu_{1i}^{\,} e^{\sigma t} + \nu_{2i}^\ast e^{\sigma^\ast t}
\qquad (i=1,2,3)\enspace
\end{align}
which allows to determine the growth rate $\sigma(k, \ldots)$ of those
perturbations. Marginal stability, i.e.  ${\rm Re }[\sigma(k, \ldots)]=0$,
then yields the stability boundary as a function of the control
parameter $\eta$. 
As an example, if one considers the stability of stripes 
%If the stability of stripes 
in a $2:1$ resonant case,
%would be of interest,
the amplitude $B=0$ and the marginal stability 
condition leads  to a polynomial in $\eta$
\begin{align}
\left(\bar\rho-1\right)\eta^2 &+\left[2\bar\rho\bar{G}_1\left(\bar\rho-1\right)-\bar\delta^2
\right]\eta\nonumber\\
&-\bar{G}_1\left(\bar\delta^2-\bar\rho^2\bar{G}_1\right)=0\, ,   
\end{align}
whose solutions are 
\begin{align}
\label{stabe1m}
\eta_{1,2}=&\frac{2\bar \rho \bar G_1\left(1 -\bar \rho\right)+
\bar \delta^2}{2\left(1-\bar \rho\right)^2}
\nonumber\\
&\pm\frac{\bar \delta}{2\left(\bar \gamma-\bar \rho\right)^2}\sqrt{\bar \delta^2+4\bar G_1\left(1-\bar \rho\right)}\, .
\end{align}
Consequently single mode solutions (stripes) are linear unstable within the
region bounded by those values of $\eta$ and linear stable in the complementary
area.
%%%%%%%%%%%%%%%%%%%%%%%%%%%%%%%%%%%%%%%%%%%%%%%%%%%%%%%
\section{Nonlinear solutions of the modulated amplitude equations}
\label{nonlinsol}
%%%%%%%%%%%%%%%%%%%%%%%%%%%%%%%%%%%%%%%%%%%%%%%%%%%%%%%%
In this section the bifurcation--scenarios from the homogeneous basic
state and thus the weakly nonlinear solutions of  Eqs.~(\ref{amplihexs}) are
discussed for three different cases: first of all we consider
an unmodulated system ($G_1=G_2=0$), then we study the 
modulated cases already taken into account in Sec.~\ref{ampleq}, namely
either $G_1 \not =0$ and $ G_2=0$ or  $G_1=0 $ and $ G_2\not =0$.
%The scenarios of the bifurcations  from the homogeneous basic
%state and therefore the nonlinear solutions of  Eqs.~(\ref{amplihexs}) 
%are determined in this section for three different cases: 
%Without modulations  ($G_1=G_2=0$) 
%and for two different modulations either  
%with $G_1 \not =0$ and $ G_2=0$ or 
%with $G_1=0 $ and $ G_2\not =0$. 
Without modulations one has either a single mode solution with
modulus $ |\bar A_1| =A$ and $|\bar A_2|=|\bar A_3|=0$, or hexagons corresponding to a 
three mode solution with coinciding moduli $ |\bar A_1| = |\bar A_2|=|\bar A_3|$. In case of
a finite modulation amplitude
%For the remaining cases of finite modulation amplitudes 
single mode solutions as well as three mode solutions with amplitudes satisfying 
$|\bar A_1|\not = |\bar A_2|=|\bar A_3|$ may be encountered. By means of the 
determined nonlinear solutions the functional value of the latter ones may be
calculated through Eq.~(\ref{pothexa}), which gives rise to the intervals
wherein either the one mode or the three mode solution is preferred.   

%%%%%%%%%%%%%%%%%%%%%%%%%%%%%%%%%%%%%%%%%%%%%%%%%%%%%%%%%%
\subsection{Solutions of the unforced amplitude equations}
%%%%%%%%%%%%%%%%%%%%%%%%%%%%%%%%%%%%%%%%%%%%%%%%%%%%%%%%

Without modulations Eqs.~\eqref{amplihexs} have either a single mode solution
$|\bar A|=\sqrt{\eta}$ or a three mode solution with equal moduli, i.e. with
amplitudes fulfilling $ |\bar A_1|=|\bar A_2|=|\bar A_3|=A$. Assuming a 
relative phase shift $\theta_j$ for each linear mode contributing to the three
mode solution
\begin{align}
\bar A_j = A \, e^{i\Theta_j}
\end{align}
one obtains a nonlinear equation for the common modulus $A$
\begin{eqnarray}
\label{noneqA}
0= \eta  A +  \bar \delta e^{i\Theta}  A^2 - (1 + 2 \bar\rho) A^3\,,
\end{eqnarray}
with $\bar \delta =\pm 1$ and 
the  sum of the phase angles $\Theta = \theta_1+\theta_2+\theta_3$.
Eq.~\eqref{noneqA} has two real solutions
\begin{equation}
\label{hexamp}
A_{\pm} = \frac{1}{2(1 +2 \bar\rho)} \left[ \bar \delta  \pm
\sqrt{ \bar \delta^2  + 4 \eta \left(1 + 2 \bar\rho \right)\,\,} \right]\,,
\end{equation}
where $A_+$ corresponds to the larger one of the two amplitudes with
$\bar \delta =1$, while $A_-$ is obtained for $\bar \delta=-1$. If 
$\bar\delta =1$ the phase angle is $\Theta=0$ and one expects 
{\it regular hexagons} whereas $\bar \delta =-1$ yields a phase angle
$\Theta=\pi$ corresponding to {\it inverse hexagons}. Comparing the 
functional given in Eq.~\eqref{pothexa} of those solutions one may 
ascertain: In case $\bar \delta=1$ regular hexagons are preferred as
${\mathcal F}_H^+ < {\mathcal F}_H^-$ and if $\bar \delta=-1$ inverse
hexagons are most likely to occur because their functional value is the lowest one,
i.e. ${\mathcal F}_H^-< {\mathcal F}_H^+$.
%Stability of hexagon (siehe Markus Manuscript).

%%%%%%%%%%%%%%%%%%%%%%%%%%%%%%%%%%%%%%%%%%%%%%%%%%%%
\subsection{Solutions of the forced amplitude equations for  
 $k_m=2k_c$ and $G_1\not =0,~ G_2=0$.}
\label{case12}
%%%%%%%%%%%%%%%%%%%%%%%%%%%%%%%%%%%%%%%%%%%%%%%%%%%%%
In the case of a $2:1$ resonance with a finite modulation amplitude  $G_1$
two of the three amplitudes coincide $A_{2,3}=B e^{i\theta_{2,3}}$, the
third one being $A_1=A e^{i\theta_1}$. Thus the  remaining  two amplitudes
$A$ and $B$ are determined by the following two  nonlinear equations
\begin{subequations}
\label{noncase1s}
\begin{align}
\label{nonc1As}
& \left( \eta -   A^2  
-2 \bar \rho   B^2\right)  A
+  \bar G_1   A  +  \bar \delta e^{i\Theta}   B^{ 2}=0\,,\\
\label{nonc1Bs}
& \left(\eta - (1+\bar \rho)   B ^2  - 
\bar \rho   A^2\right) B
   +  \bar \delta e^{i\Theta}  A  B
=0\, ,
\end{align}
\end{subequations}
with $\Theta=\theta_1+\theta_2+\theta_3$. The single mode solution 
generated by those coupled equations is given by 
\begin{align}
\label{stripesol}
A= \sqrt{\eta + \bar G_1}\, ,\qquad B=0
\end{align}
and bifurcates supercritcally at the threshold $\varepsilon =-G_1$ 
or $\eta = -\bar G_1$ in rescaled units.
A linear stability analysis of the single mode solution
according to the scheme described in 
Sec.~\ref{linstabs} shows that the single mode
is linear unstable in the region delimited by the two
values of $\eta$
\begin{align}
\label{stabe1m}
\eta_{1,2}=&\frac{2\bar \rho \bar G_1\left(1 -\bar \rho\right)+
\bar \delta^2}{2\left(1-\bar \rho\right)^2}
\nonumber\\
&\pm\frac{\bar \delta}{2\left(\bar \gamma-\bar \rho\right)^2}\sqrt{\bar \delta^2+4\bar G_1\left(1-\bar \rho\right)}\, .
\end{align}

\begin {figure}[hbt]
\psfrag{gamm_gr_null}{\Large$\gamma>0$}
 %\vspace{4cm}
\psfrag{eps}{\large$\eta$}
\psfrag{(a)}{\small$(a)$}
\psfrag{(b)}{\small$(b)$}
\psfrag{(c)}{\small(c)}
\psfrag{(d)}{\small(d)}
\psfrag{B}{\small$B$}
\hspace{-5mm}\psfrag{AB}{\small$A,B$}
\begin{center}
\includegraphics [width=0.9\columnwidth] {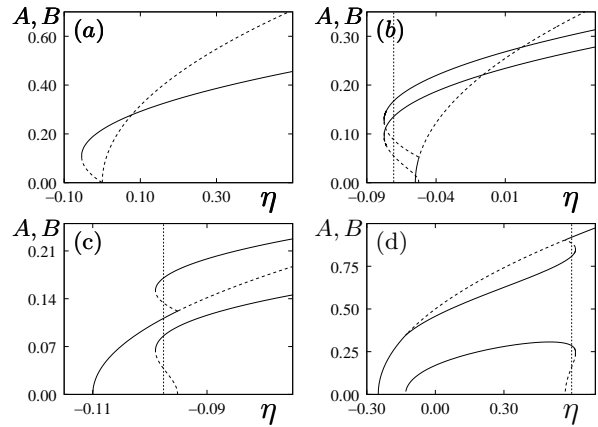}
\end{center}
\caption{The amplitude of  supercritically   bifurcating stripes and the
subcritically bifurcating hexagons with three degenerated amplitudes 
 are shown in (a) as a function of $\eta$
for the unmodulated case.
In each case the solid lines mark the linear stable part of 
each branch and the dashed parts the  unstable ones.
For finite modulation 
amplitudes $\bar G_1$ the three mode solution with $A$
always corresponding to the upper branch and $B$ to the lower
branch, as well as the amplitude of a single mode solution 
are plotted as a function of the control parameter $\eta$ 
for three different values of the modulation amplitude 
$G_1=0.05$ (b), $G_1=0.11$ (c) and $G_1=0.25$ (d)
whereas the nonlinear coefficient $\bar\rho$ is fixed at $\bar\rho=1.8$
and $\bar\delta =1$.
 Each vertical
dotted line reflects the $\eta$-value where the values of the functional 
${\cal F}$  of the  one and the three mode solution coincide.
}
\label{bifg1}
\end{figure}

The  three mode solution with $\bar \delta ~e^{i\Theta}=1$ is generally
preferred and the appendant amplitudes $A$ and $B$ may be determined by
solving Eqs.~(\ref{noncase1s}) with a standard computer algebra program.
Instead of presenting the rather complex analytical solutions we opt for
the characteristic bifurcation diagrams based on four different parameter
sets in Fig.~\ref{bifg1}, which ought to cover the typical bifurcation 
scenarios encountered. In Fig.~\ref{bifg1} the linear stable intervals
of each branch are visualized by solid lines, whereas unstable ones are
represented by dashed lines.  
% *******************************************************************
%   Figure
% *******************************************************************
%
%%%%%%%%%%%%%%%%%%%%%%%%%%%%%%%%%%%%%%%%%%%%%%%%%%%%%%%%%%%%%%%%%%%%%%%%%%%%%%%%%%5

Part (a) of  Fig.~\ref{bifg1} shows the well known bifurcation scenario
of the unmodulated case $(\bar{G}_1=0)$ for supercritcally bifurcating stripes
and subcritically bifurcating hexagonal solutions \cite{Ciliberto:90.1}.
With increasing values of the modulation amplitude $\bar G_1$ the degeneracy  
%that the three modes of the hexagonal solution show in part (a), is abolished 
of the three mode solution vanishes and only the onset of the single mode
solution starts at 
$\eta_c=-\bar{G}_1$. This removal of degeneracy is already best reflected in 
Fig.~\ref{bifg1}(b), wherein $\bar{G}_1=0.05$: the subcritical branches of 
the hexagonal solution, $A$ corresponding to the larger amplitude and $B$ to
the smaller one, do not overlap anymore. For finite values of $\bar G_1$ 
the upper hexagon-branch, i.e. the $A$-branch, always emerges from the
single mode solution, while the $B$-branch which is the lower hexagon-branch
bifurcates from zero with decreasing values $\eta$. If $\bar G_1$
is small enough the bifurcation of the three-mode solution is subcritical
as depicted in Fig.~\ref{bifg1} (b) and  Fig.~\ref{bifg1}(c), whereas for large
modulation amplitudes this bifurcation is shifted to smaller $\eta$-values while
becoming supercritical as sketched in Fig.~\ref{bifg1} (d).
% ***********************************************************
%   Figure
% ***********************************************************
%
\begin {figure}[hbt]
%\vspace{4cm}
\psfrag{eps}{\large$\eta$}
\psfrag{AB}{\hspace{-0.1cm}\small$A,B$}
\psfrag{(a)}{\small(a)}
\psfrag{(b)}{\small(b)}
\begin{center}
\includegraphics [width=0.75\columnwidth] {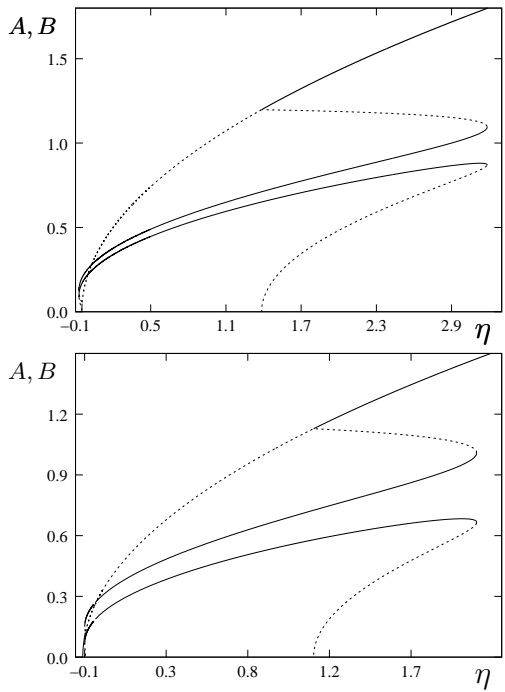}
%\includegraphics [width=0.75\columnwidth] {./figures/Hex005F1_ZoomOut.eps}\\
%\vspace{2mm}
%\includegraphics [width=0.75\columnwidth] {./figures/Hex011F1_ZoomOut.eps}
\end{center}
\caption{The amplitudes $A(\eta)$ and $B(\eta)$ 
are plotted for 
$\bar G_1=0.05$ and $\bar G_1=0.11$ (already shown in Fig.~\ref{bifg1})
 over a broader interval
of $\eta$ which reveals
that the upper bifurcation point of the three mode
solution moves for increasing modulation amplitudes 
$\bar{G}_1$ to decreasing values of $\eta$. As in
Fig.~\ref{bifg1} $\bar\rho =1.8$ and $\bar\delta =1$.
}
\label{bifg1L}
\end{figure}
%%%%%%%%%%%%%%%%%%%%%%%%%%%%%%%%%%%%%%%%%%%%%%%%%%%%%%%%%%%%%%%%%%%%%%%%%%%%%%%%%%5

\begin {figure}[hbt]
 %\vspace{4cm}
\psfrag{eps}{\vspace{0.2cm}\small$\bar{G}_1$}
\psfrag{AB}{\hspace{0.55cm}\large$\eta$}
\psfrag{(a)}{\small(a)}
\psfrag{(b)}{\small(b)}
\begin{center}
\includegraphics [width=0.85\columnwidth] {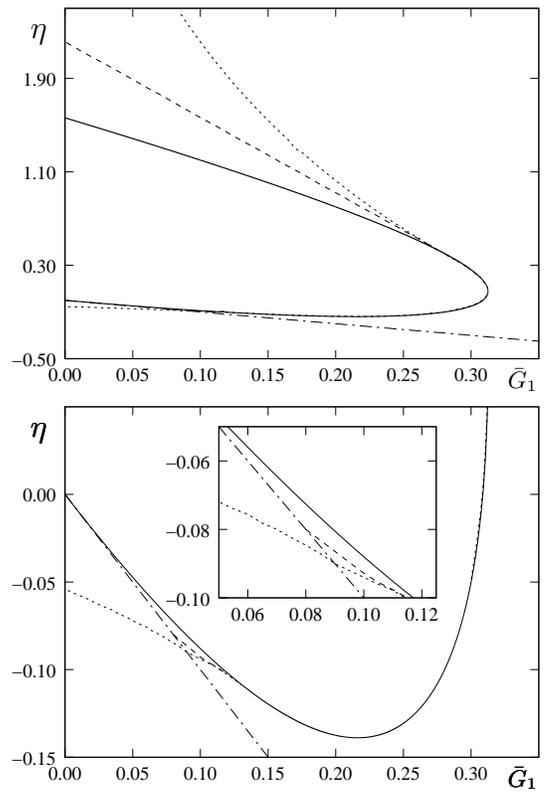}
%\includegraphics [width=0.85\columnwidth] {./figures/StabFall1.eps}\\
%\vspace{0.2cm}
%\includegraphics [width=0.85\columnwidth] {./figures/StabFall1Zoom.eps}
\end{center}
\caption{Between the upper and lower dotted lines
(the lower one overlapping partially with the solid line)
lies the domain of existence of three mode solutions. 
The functionals of single and three mode solutions 
coincide along the dashed line (which overlaps 
once again partially on its lower branch with the
solid one) and the region lying in between these 
lines is the one where three mode solutions are 
favored. Single mode solutions are linear unstable 
inside the area delimited by the solid line whereas
linear stable outside and the dashed--dotted line 
corresponds to the threshold $\eta_c=-\bar{G}_1$ of
the single mode solution. Part (b) of the figure holds an 
enlargement of the small $\eta$ range in order to
distinguish properly the different curves in the 
lower half plane of part (a). Further parameters are
$\bar\rho =1.8$ and $\bar\delta=1$. 
}
\label{bifstaba}
\end{figure}

It turns out that for intermediate values of $\bar{G}_1$, such as
those used in Fig.~\ref{bifg1}(b) and  Fig.~\ref{bifg1}(c), the three
mode solution bifurcates subcritically from the single mode solution
not only for small values of the control parameter $\eta$ but also for
large values of $\eta$ as it may be concluded from Fig.~\ref{bifg1L}.
For large values of $\bar{G}_1$, e.g. $\bar{G}_1=0.25$ in Fig.~\ref{bifg1}(d),
the lower bifurcation point becomes supercritical while the upper one
nevertheless remains subcritical at first. By comparing the bifurcation diagrams
shown in Fig.~\ref{bifg1L} and Fig.~\ref{bifg1}(d) it can easily be seen
that the existence range of the three mode solution shrinks with increasing
values of the modulation amplitude. Simultaneously the hysteresis becomes
less pronounced with increasing values of $\bar G_1$ changing the 
subcritical bifurcation to a supercritical one at small $\eta$-values 
as well as the upper bifurcation point from subcritical to supercritical
for even larger values of $\bar{G}_1$. This latter point can be nicely seen in Fig.~\ref{bifstaba} which shows that beyond some critical value $\bar{G}_{1c}$
of the modulation amplitude the three mode solution ceases to exist.
Since both bifurcations from the single to the three mode solution are already
supercritical before their domain of existence shrinks to zero, the critical
value $\bar G_{1c}$ may be determined using the condition that the two 
$\eta$--values given by  Eq.~(\ref{stabe1m}) coincide, i.e. $\eta_1=\eta_2$. 
Hence for modulation amplitudes $\bar{G}_1$ greater than the critical 
value
\begin{align}
\label{stabe2m}
\bar{G}_{1c}= \frac{1}{4(\bar{\rho}-1)}
\end{align}
single mode solutions are stable for all $\eta$ and three mode solutions
do not exist any more. Since Fig.~\ref{figrho} evinces that in most cases
$\bar \rho$ takes values larger than one, $\bar G_{1c}$ should always be a
positive parameter.
% ***********************************************************
%   Figure
% ***********************************************************
%
%%%%%%%%%%%%%%%%%%%%%%%%%%%%%%%%%%%%%%%%%%%%%%%%%%%%%%%%%%%%%%%%%%%%%%%%%%%%

The domain of existence of three mode solutions, the stability
range of single mode solutions as well as the points of
coincidence of the corresponding functionals (vertical dotted
lines) are represented in Fig.~\ref{bifg1}(b)-(d) and 
Fig.~\ref{bifg1L} for three different values of $\bar{G}_1$.
These relevant properties are recapitulated as a function of
the modulation amplitude $\bar G_1$ in Fig.~\ref{bifstaba}.
Within the region bounded by the dotted lines three mode 
solutions exist. Along the dashed line the functional values
of one and three mode solutions match and the area lying in
between the upper and lower solid lines is the portion of the
$\eta$-$\bar{G}_1$-plane wherein one mode solutions are linear
unstable. As predicted by Eq.~(\ref{stabe2m}) for $\bar\rho=1.8$
three mode solutions do not exist for modulation amplitudes 
greater than $\bar{G}_1 =0.3125$. Before this critical value is
reached the dotted and the solid lines are overlapping which
means that the upper bifurcation of three mode solutions, i.e.
the one occurring at large $\eta$-values, has become supercritical 
as it has already been mentioned above. 

%
%
% *************************************************************
%   Figure
% *************************************************************
%
\begin{figure}[hbt]
\begin{center}
%
%\vspace{-5mm}
\psfrag{d}{\small $d$}
\psfrag{alp}{\small $\alpha$}
\psfrag{del}{\small $\delta$}
\psfrag{gam}{\small $\gamma$}
\psfrag{rho}{\small $\rho$}
\includegraphics [width=0.95\columnwidth] {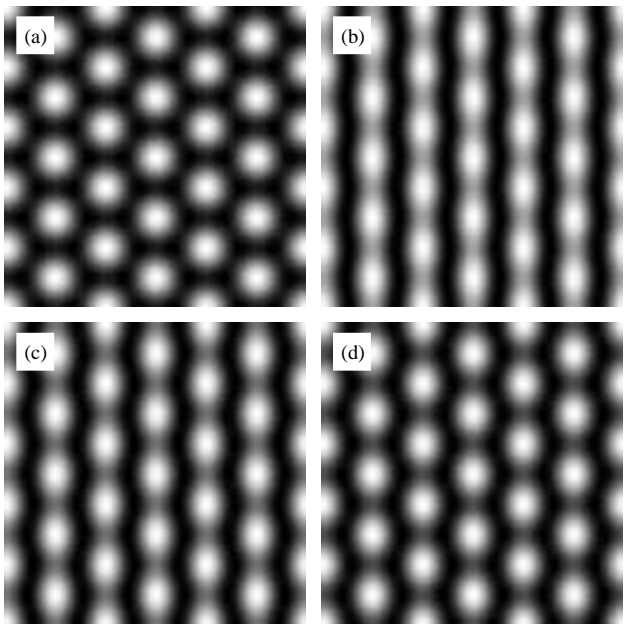}
%\includegraphics [width=0.48\columnwidth] {./figures/muster01.eps}
%\includegraphics [width=0.48\columnwidth] {./figures/muster02.eps}
%\\[2mm]
%\includegraphics [width=0.48\columnwidth] {./figures/muster05.eps}
%\includegraphics [width=0.48\columnwidth] {./figures/muster06.eps}
\end{center}
%\vspace{-0.7cm}
%
\caption{\label{planeform}
The spatial variation of one term contributing to   
Eq.~(\ref{hexansatz}) is shown for hexagons in part (a) and
for distorted hexagons in (b), (c) and (d). In part (b) a
distorted hexagon is depicted for the parameters used in Fig.~\ref{bifg1}(b),
namely $\bar G_1=0.05$ and $\eta=-0.0492$ with the amplitudes being   
$A=0.0551$ and $B=0.0124$. 
In the remaining two parts the modulation amplitude is
$\bar G_1=0.11$, whereas the control parameter is $\eta=-0.099$ in (c)
(with $A=0.15$ and $B=0.06$) 
 and  $\eta=-0.09$ in
part (d) (with  $A=0.198$ and $B=0.116$).
}
\end{figure}
%%%%%%%%%%%%%%%%%%%%%%%%%%%%%%%%%%%%%%%%%%%%%%%%%%%%%%%%%%%%%%%%%%%%%%%%%%%%%%%%%%%%%%%%%%
According to Eq.~(\ref{hexansatz}) different amplitudes $A$ and $B$ generate  
dissimilar spatial patterns as illustrated in Fig.~\ref{planeform}
for four distinct parameter sets. Fig.~\ref{planeform}(a) shows
regular hexagons whereas in Fig.~\ref{planeform}(b)-(d) distorted
hexagons can be seen for  various amplitude ratios $A/B$ corresponding
to the parameters used in Fig.~\ref{bifg1} (b) and (c).

%%%%%%%%%%%%%%%%%%%%%%%%%%%%%%%%%%%%%%%%%%%%%%%%%%%%
\subsection{
Solutions of the forced amplitude equations for  
 $k_m=k_c$ and $G_1=0,~ G_2\not=0$.}
\label{case11}
\begin {figure}[hbt]
\psfrag{eps}{\large$\eta$}
\psfrag{AB}{\small$A,B$}
\psfrag{(a)}{\small$(a)$}
\psfrag{(b)}{\small$(b)$}
\psfrag{(c)}{\small(c)}
\psfrag{(d)}{\small(d)}
\begin{center}
\includegraphics [width=0.9\columnwidth] {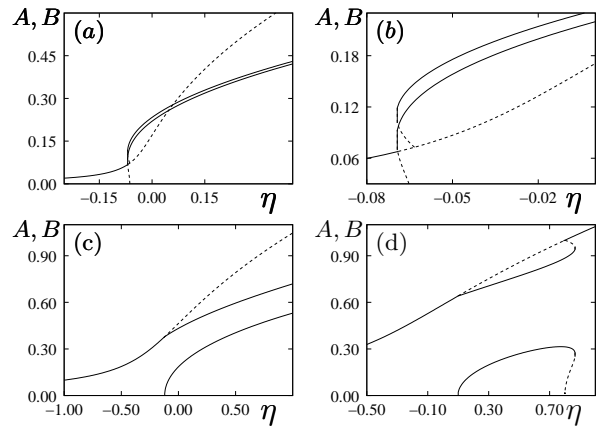}
\end{center}
\caption{The amplitudes of a three mode solution with $A$ 
always corresponding to the upper branch and $B$ to the lower
one, as well as the amplitude of the imperfect bifurcating 
single mode solution ($B=0$) are plotted as a function of the
control parameter $\eta$. Three different values of the 
modulation amplitude are taken into account: $\bar{G}_2=0.05$ (a),
where (b) is an enlargement of (a) over a smaller $\eta$-range,
$\bar G_2=0.1$ (c) and $\bar G_2=0.2$ (d). Solid lines mark linear 
stable solutions and dashed lines unstable ones. Furthermore 
$\bar\rho =1.7$ and $\bar\delta =1$.
}
\label{bifg2}
\end{figure}

%%%%%%%%%%%%%%%%%%%%%%%%%%%%%%%%%%%%%%%%%%%%%%%%%%%%%
As in the previous case two amplitudes of the three mode solution
coincide, i.e. $A_{2,3}=B e^{i\theta_{2,3}}$, whereas the third one is
given by $A_1=A e^{i\theta_1}$. The two remaining moduli $A$ and $B$ 
are determined by the following nonlinear equations
\begin{subequations}
\label{noncase2s}
\begin{align}
\label{nonc2As}
& \left( \eta - \mid \bar A\mid^2  
-2 \bar \rho \mid \bar B\mid^2\right) \bar A
+  \bar G_2    +  e^{i\Theta}  \bar B^{\ast 2}=0\,,\\
\label{nonc2Bs}
& \left(\eta - (1+\bar \rho) \mid \bar B\mid ^2  - 
\bar \rho \mid \bar A\mid^2\right)\bar B
   +  e^{i\Theta} \bar A^\ast \bar B^\ast 
=0\,.
\end{align}
\end{subequations}
If $B=0$ and $\bar G_2 \not =0$ the single mode solution is
determined by a third order polynomial in $A$. In contrast to
the case discussed in Sec.~\ref{case12} the bifurcation is now an 
imperfect one as can be seen in Fig.~\ref{bifg2}. Despite this
imperfect bifurcation one may observe the following similarities
to the previous paragraph  wherein the $2:1$ resonance was studied: First of 
all the degeneracy of three mode solutions encountered for 
$\bar{G}_2=0$ vanishes once again for finite values of the 
modulation amplitude $\bar G_2$. Secondly, as for the $2:1$ 
resonance, $A$ corresponding to the largest amplitude of the three mode solution 
bifurcates for small values of $\bar{G}_2$ from the top of
the single mode solution, whereas the small amplitude $B$
bifurcates from zero.  
%
% *******************************************************************
%   Figure
% *******************************************************************
%
%%%%%%%%%%%%%%%%%%%%%%%%%%%%%%%%%%%%%%%%%%%%%%%%%%%%%%%%%%%%%%%%%%%%%%%%%%%%%%%%%%

% *******************************************************************
%   Figure
% *******************************************************************
%
\begin {figure}[hbt]
\psfrag{eps}{\small$\bar{G}_2$}
\psfrag{AB}{\hspace{0.55cm}\large$\eta$}
\includegraphics [width=0.9\columnwidth] {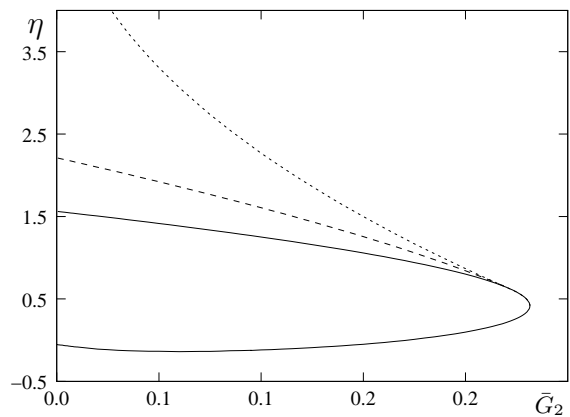}
\caption{In between the dotted line and the lower solid line 
(coinciding in the lower half--plane with the dashed line)
three mode solutions exist. Along the dashed line, which 
overlaps on its lower branch with lower branch of the
solid line, the values of the functionals of single and
three mode solutions match. Thus within the area bounded by 
the dashed line three mode solutions are favored. Single mode
solutions are linear stable in the complementary part of the 
region enclosed by the solid line. The nonlinear coefficient
is assumed to be $\bar\rho =1.7$ and $\bar\delta =1$.
}
\label{hexG}
\end{figure}

With increasing values of $\bar{G}_2$ the bifurcation of the three mode 
solution changes once more at its lower bifurcation point from a 
subcritical one in Fig.~\ref{bifg2}(a) to a supercritical one in 
Fig.~\ref{bifg2}(c) or in Fig.~\ref{bifg2}(d). The bifurcation point
located at large values of $\eta$ is not shown in Fig.~\ref{bifg2}(a)
and Fig.~\ref{bifg2}(c) but is subcritical as the one depicted in 
Fig.~\ref{bifg1L} for a $2:1$ resonance. Thirdly the hysteresis described
by the three mode solution becomes once again less pronounced with
increasing values of $\bar{G}_2$ and turns out to be supercritical for
large modulation amplitudes. Last but not least the three mode solution
ceases to exist beyond some critical value of the modulation amplitude 
$\bar{G}_2$ as it was noticed as well for a $2:1$ resonance.

The properties explained in the previous paragraph are recapitulated in 
Fig.~\ref{hexG}, where three mode solutions exist within the region 
bounded by the dotted lines, overlapping in the lower half--plane with
the solid line. Along the dashed line the functionals ${\cal F}$ of the
single and three mode solution match, meaning that
the three mode solution is preferred in between the dashed  lines coinciding 
once more with the solid line for small values of the control parameter $\eta$.
This overlapping of lines is due to the fact that the lower bifurcation point
is supercritical for small $\bar{G}_2$ and that the extension of the hysteresis
shrinks with increasing modulation amplitudes. Through a linear stability 
analysis, similar to the one described in Sec.~\ref{linstabs}, one may conclude
that single mode solutions are linear stable only outside of the area delimited 
by the solid line.  

%%%%%%%%%%%%%%%%%%%%%%%%%%%%%%%%%%%%%%%%%%
\section{Numerical results}\label{Simul}
%%%%%%%%%%%%%%%%%%%%%%%%%%%%%%%%%%%%%%%%%%

\begin {figure}[htb]
\centering
\psfrag{gamm_gr_null}{\Large$\gamma>0$}
 %\vspace{4cm}
\psfrag{eta}{\Large$\eta$}
\psfrag{A}{\large$A$}
\includegraphics [width=0.9\columnwidth] {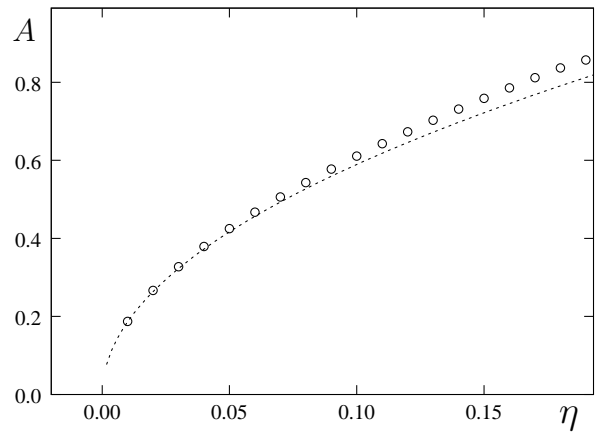}
\caption{
The amplitude of a stripe solution obtained by numerical simulations 
(circles) of the microscopic model given by Eqs.~(\ref{LE}) is compared
with the analytical solution $A =\sqrt{\eta/\gamma}$. The parameters used
are $a=16, \sigma=301, d=1.07, c=0.6$ which yield the nonlinear coefficient $\gamma=0.288$.
For the sake of simplicity no modulations were taken into account.
}
\label{simstripe}
\end{figure}

A major problem always encountered when dealing with amplitude equations is that
their validity range around the threshold remains a priori unpredictable. 
Therefore we have determined the amplitude of the stripe solution
by numerical simulations  of the basic Eqs.~(\ref{LE}) 
without considering modulations as a function of the relative distance
$\eta$ from threshold 
and compared these numerical results with the analytical solution 
$A=\sqrt{\eta/\gamma}$ in Fig.~\ref{simstripe}. Up to $\eta \sim 0.2$ we find a 
fairly good agreement between analytical and numerical results, which gives a 
reasonable estimate of the quantitative validity range of the preceding results 
obtained in terms  of amplitude equations.
% *******************************************************************
%   Figure
% *******************************************************************
%
%\begin {figure}[hbt]
%\psfrag{gamm_gr_null}{\Large$\gamma>0$}
% \vspace{4cm}
%\includegraphics [width=8.cm] {./figures/hex13_gamma0.eps}
%
%
%\caption{ In a) the  amplitude $A$ 
%is shown as function of $\eta$ for the stripe solution
%for $\bar G_2=0.05,~0.2,~1$ (oder \"ahnlich). In b) for the three mode solution 
%the amplitude $A$ und in c) the amplitude $B$ is shown
%for same values of $\bar G_2$ and $\bar \rho=2$. 
%}
%
%\label{stabg3}
%%
%\end{figure}
% *******************************************************************
%

%%%%%%%%%%%%%%%%%%%%%%%%%%%%%%%%%%%%%%%%%%%%%%%%%%%%%%%%%%%%%%%%%%
\section{Summary and conclusions}%%%%%
\label{conclu}
%%%%%%%%%%%%%%%%%%%%%%%%%%%%%%%%%%%%%%%%%%%%%%%%%%%%%%%%%%%%%%%%%%

The generic amplitude equations (\ref{amplihex}) are suggested  
for a forced two-dimensional system with a broken up-down symmetry 
and  discussed throughout this work. Additionally 
they are derived for the first time explicitly from a 
model for chemical reactions,
 in the present case the Lengyel-Epstein model. They are apt to
describe how spatial periodic modulations interfere with the competition
between stripes and hexagonal patterns. Our interest was particularly 
focused on a $1:1$ and a $2:1$ resonance between the modulation wave
number $k_m$ and the wave number $k_c$ of the respective pattern.
Compared to the unmodulated case the bifurcation--scenarios of stripes 
and hexagonal patterns change strongly in the presence of 
modulations. It has been shown that three mode solutions, i.e. 
hexagons in the unmodulated case and distorted hexagons otherwise,
are suppressed for large modulation
amplitudes. Furthermore evidence was given that depending on the strength
of the modulation amplitude those three mode solutions bifurcate either sub- or
supercritically. The bifucation diagrams presented in 
this work may be confirmed by experiments on chemical
reactions with a spatially modulated illumination, where in contrast to  Refs.~\cite{Epstein:2001.1,Epstein:2003.1}  
the ratio between the wave number of the illumination 
and the wave number of the Turing pattern should be chosen to be
$2:1$ or  $1:1$.

Instead of considering a stationary forcing the effect of traveling stripe
forcing on the competition between hexagons and stripes has recently been
investigated for a $1:1$ resonance in a one-dimensional system \cite{Sagues:2003.1}.  
Given the coupled amplitude equations (\ref{amplihex}) the results 
established in Ref.~\cite{Sagues:2003.1} may be directly interpreted within the scope of 
the microscopic Lengyel-Epstein model. The same allegation holds for the $1:1$
resonance analyzed in a two--dimensional system in Ref.~\cite{Kramer:2004.1}.
Although the bifurcation--scenarios of the stationary $1:1$ and $2:1$ resonance we studied 
in the present work share some similarities, there remains the open question 
%which might be of some interest:
how the bifurcation diagrams presented in Refs.~\cite{Sagues:2003.1,Kramer:2004.1}
are modified if one has to deal with a $2:1$ resonance instead of a $1:1$ resonance.

We mainly focused on the spatially periodic forcing of Turing patterns. The 
Lengyel--Epstein model exhibits furthermore a Hopf-bifurcation \cite{Epstein:1993.1} similar to
those found in other chemical reaction schemes. Instead of a temporal resonant 
forcing of an oscillatory but spatially homogeneous chemical reaction \cite{Swinney:2000.1,Swinney:2000.2} one
may induce such an oscillatory reaction by a spatially
periodic or a traveling stripe modulation. This leads to a revealing
amplitude equation, which has been recently deduced from 
the Lengyel-Epstein model and exhibits rich transition scenarios
between solutions that are either spatially harmonic or subharmonic with respect
to the external modulation \cite{Hammele:2004.4}.
%\renewcommand{\baselinestretch}{1.0}

%\begin{acknowledgement}
This work results mainly from a summer project for students. 
%\end{acknowledgement}

%
\bibliographystyle {prsty}
\end{document}